\documentclass[10pt]{article}
\usepackage{latexsym,graphicx}
\newcommand{\be}{\begin{equation}}
\newcommand{\ee}{\end{equation}}
\def\n{\noindent}
\catcode `\@=11
\catcode `\@=12
\begin{document}
\begin{center}
\large{\bf {Plane-symmetric inhomogeneous  magnetized viscous fluid universe with a variable $\Lambda$}} \\
\vspace{10mm}
\normalsize{ANIRUDH PRADHAN $^1$, PURNIMA PANDEY $^2$} \\
\normalsize{$^{1, 2}$ \it {Department of Mathematics, Hindu Post-graduate College, 
 Zamania-232 331, Ghazipur, India}} \\
\normalsize{$^1$ E-mail : pradhan@iucaa.ernet.in, acpradhan@yahoo.com} \\
\normalsize{$^2$ E-mail : purnima\_pandey2001@yahoo.com} \\
%\normalsize{$^2$ \it {}\\
\end{center}
\vspace{10mm}
%\date{}
%\maketitle
\begin{abstract} 
The behavior of magnetic field in plane symmetric inhomogeneous cosmological models for
bulk viscous distribution is investigated. The coefficient of bulk viscosity is assumed 
to be a power function of mass density $(\xi = \xi_{0}\rho^{n})$. The values of cosmological 
constant for these models are found to be small and positive which are supported by the 
results from recent supernovae Ia observations. Some physical and geometric aspects of the 
models are also discussed.
\end{abstract}
\smallskip
\n PACS: 98.80.-k, 98.80.Es \\
\n Key words : cosmology, inhomogeneous models, electro-magnetic field, bulk viscous models\\

\section{Introduction}     %\section*{Introduction}

The standard Friedmann-Robertson-Walker (FRW) cosmological model prescribes a homogeneous
and an isotropic distribution for its matter in the description of the present state of 
the universe. At the present state of evolution, the universe is spherically symmetric and
the matter distribution in the universe is on the whole isotropic and homogeneous. But in
early stages of evolution, it could have not had such a smoothed picture. Close to the 
big bang singularity, neither the assumption of spherical symmetry nor that of isotropy can
be strictly valid. So we consider plane-symmetric, which is less restrictive than spherical
symmetry and can provide an avenue to study inhomogeneities. Inhomogeneous cosmological 
models play an important role in understanding some essential features of the universe 
such as the formation of galaxies during the early stages of evolution and process of 
homogenization. The early attempts at the construction of such models have done by 
Tolman\cite{ref1} and Bondi\cite{ref2} who considered spherically 
symmetric models. Inhomogeneous plane-symmetric models were considered by 
Taub\cite{ref3,ref4} and later by Tomimura\cite{ref5}, Szekeres\cite{ref6}, Collins and 
Szafron\cite{ref7}, Szafron and Collins\cite{ref8}. Recently, Senovilla\cite{ref9} obtained 
a new class of exact solutions of Einstein's equation without big bang singularity, 
representing a cylindrically symmetric, inhomogeneous cosmological model filled with 
perfect fluid which is smooth and regular everywhere satisfying energy and causality 
conditions. Later, Ruis and Senovilla\cite{ref10} have separated out a fairly large 
class of singularity free models through a comprehensive study of general cylindrically
symmetric metric with separable function of $r$ and $t$ as metric coefficients.
Dadhich {\it et al.}\cite{ref11} have established a link between the FRW model and the 
singularity free family by deducing the latter through a natural and simple
in-homogenization and anisotropization of the former. Recently, Patel {\it et al.}\cite{ref12}
presented a general class of inhomogeneous cosmological models filled with 
non-thermalized perfect fluid by assuming that the background space-time admits two space-like
commuting killing vectors and has separable metric coefficients. Bali and Tyagi\cite{ref13}
obtained a plane-symmetric inhomogeneous cosmological models of perfect fluid distribution
with electro-magnetic field. Recently, Pradhan {\it et al.}\cite{ref14} have investigated
a plane-symmetric inhomogeneous viscous fluid cosmological models with electro-magnetic
field. \\    

Models with a relic cosmological constant $\Lambda$ have received considerable 
attention recently among researchers for various reasons 
(see Refs.{\cite{ref15}}$-${\cite{ref19}} and references therein). Some of the 
recent discussions on the cosmological constant ``problem'' and consequence on cosmology 
with a time-varying cosmological constant by Ratra and Peebles\cite{ref20}, 
Dolgov\cite{ref21}$-$\cite{ref23} and Sahni and Starobinsky\cite{ref24} have
pointed out that in the absence of any interaction with matter or radiation, the 
cosmological constant remains a ``constant''. However, in the presence of
interactions with matter or radiation, a solution of Einstein equations and the 
assumed equation of covariant conservation of stress-energy with a time-varying 
$\Lambda$ can be found. For these solutions, conservation of energy requires 
decrease in the energy density of the vacuum component to be compensated by a 
corresponding increase in the energy density of matter or radiation. Earlier 
researchers on this topic, are contained in Zeldovich\cite{ref25} 
Weinberg\cite{ref16} and Carroll, Press and Turner\cite{ref26}. Recent
observations by Perlmutter {\it et al.}\cite{ref27} and Riess {\it et al.} \cite{ref28}
strongly favour a significant and positive value of $\Lambda$. Their finding arise from 
the study of more than $50$ type Ia supernovae with redshifts in the range
$0.10 \leq z \leq 0.83$ and these suggest Friedman models with negative pressure
matter such as a cosmological constant $(\Lambda)$, domain walls or cosmic strings (Vilenkin
\cite{ref29}, Garnavich {\it et al.}\cite{ref30}). Recently, Carmeli and Kuzmenko\cite{ref31}
have shown that the cosmological relativistic theory (Behar and Carmeli\cite{ref32})
predicts the value for cosmological constant $\Lambda = 1.934\times 10^{-35} s^{-2}$.
This value of ``$\Lambda$'' is in excellent agreement with the measurements recently obtained 
by the High-Z Supernova Team and Supernova Cosmological Project (Garnavich {\it et al.}
\cite{ref30}; Perlmutter {\it et al.}\cite{ref27}; Riess {\it et al.}\cite{ref28}; Schmidt 
{\it et al.}\cite{ref31}). The main conclusion of these observations is that the expansion 
of the universe is accelerating. \\

Several ans$\ddot{a}$tz have been proposed in which the $\Lambda$ term decays 
with time (see Refs. Gasperini\cite{ref34,ref35}, Berman\cite{ref36}, 
Freese {\it et al.}\cite{ref19}, $\ddot{O}$zer and Taha\cite{ref19}, 
Peebles and Ratra\cite{ref37}, Chen and Hu\cite{ref38}, Abdussattar and Viswakarma\cite{ref39},
Gariel and Le Denmat\cite{ref40}, Pradhan {\it et al.}\cite{ref41}). Of the special interest 
is the ans$\ddot{a}$tz $\Lambda \propto S^{-2}$ (where $S$ is the scale factor of the
Robertson-Walker metric) by Chen and Wu\cite{ref38}, which has been 
considered/modified by several authors ( Abdel-Rahaman\cite{ref42}, 
Carvalho {\it et al.}\cite{ref19}, Waga\cite{ref43}, Silveira and Waga\cite{ref44},
Vishwakarma\cite{ref45}). \\

Most cosmological models assume that the matter in the universe can be described by 'dust'
(a pressure-less distribution) or at best a perfect fluid. 
However, bulk viscosity is expected to play an important role at certain stages
of expanding universe\cite{ref46}$-$\cite{ref48}. It has been shown that bulk
viscosity leads to inflationary like solution\cite{ref49}, and acts like a negative
energy field in an expanding universe\cite{ref50}. Furthermore, there are several
processes which are expected to give rise to viscous effects. These are the decoupling
of neutrinos during the radiation era and the decoupling of radiation and matter 
during the recombination era. Bulk viscosity is associated with the Grand Unification 
Theories (GUT) phase transition 
and string creation. Thus, we should consider the presence of a material distribution other
than a perfect fluid to have realistic cosmological models (see Gr\o n\cite{ref51} for a
review on cosmological models with bulk viscosity). A number of authors have discussed 
cosmological solutions with bulk viscosity in various context\cite{ref51}$-$\cite{ref54}. \\

The occurrence of magnetic fields on galactic scale is well-established 
fact today, and their importance for a variety of astrophysical phenomena
is generally acknowledged as pointed out Zeldovich {\it et al.}\cite{ref55}. Also 
Harrison\cite{ref56} has suggested that magnetic field could have a 
cosmological origin. As a natural consequences, we should include magnetic 
fields in the energy-momentum tensor of the early universe.
The choice of anisotropic cosmological models in Einstein system of field 
equations leads to the cosmological models more general than Robertson-Walker 
model\cite{ref57}. The presence of primordial magnetic fields in the early 
stages of the evolution of the universe has been discussed by several authors\cite{ref58}$-$\cite{ref67}. 
Strong magnetic fields can be created due to 
adiabatic compression in clusters of galaxies. Large-scale magnetic fields 
give rise to anisotropies in the universe. The  anisotropic pressure created 
by the magnetic fields dominates the evolution of the shear anisotropy and it 
decays slower than if the pressure was isotropic\cite{ref68,ref69}. Such 
fields can be generated at the end of an inflationary epoch\cite{ref70}$-$\cite{ref74}. 
Anisotropic magnetic field models have significant contribution 
in the evolution of galaxies and stellar objects. The model studied by Murphy\cite{ref88} 
possessed an interesting feature is that the big bang type of
singularity of infinite spacetime curvature does not occur to a finite past.
However, the relationship assumed by Murphy between the viscosity coefficient and
the matter density is not acceptable at large density. Several authors
\cite{ref75}$-$\cite{ref81} have investigated cosmological 
models with a magnetic field in different context. \\

Recently Bali {\it et al.}\cite{ref82} obtained some plane-symmetric 
inhomogeneous cosmological models for a perfect fluid distribution with 
electro-magnetic field. Motivated the situations discussed above, in this paper, 
we shall focus upon the problem of establishing a formalism for studying the 
general relativistic evolution magnetic inhomogeneities in presence of bulk 
viscous in an expanding universe. We do this by extending the work of Bali
{\it et al.}\cite{ref82} by including an electrically neutral bulk viscous fluid 
as the source of matter in the energy-momentum tensor. This paper is organized as 
follows. The metric and the field equations are presented in section 2. In section
 3 we deal with the  solution of the field equations in presence of bulk viscous 
fluid. The sections 3.1, 3.2 and 3.3  contain the two different cases (i.e. 
for $n = 0$ and $n = 1$ ) and also contain some physical aspects of these models 
respectively. Section $4$ describe some other generated  models and their physical and 
geometric properties. Finally in section $5$ concluding remarks have been given.\\ 
%%%%%%%%%%%%%%%%%%%%%%%%%%%%%%%%%%%%%% SECTION 2 %%%%%%%%%%%%%%%%%%%%%%%%%%%%%%%%%%%%%%%

\section{The metric and field  equations}
We consider the metric in the form  
\begin{equation} 
\label{eq1}  
ds^{2} = B^{2}(dx^{2} - dt^{2} + dy^{2}) + C^{2}dz^{2},
\end{equation} 
where the metric potential $B$ and $C$ are functions of $x$ and $t$. \\
The energy momentum tensor in the presence of bulk stress has the form
\begin{equation} 
\label{eq2}
T^{j}_{i} = (\rho + \bar{p})v_{i}v^{j} + \bar{p}g^{j}_{i} + E^{j}_{i},
\end{equation} 
where $E^{j}_{i}$ is the electro-magnetic field given by
\begin{equation} 
\label{eq3}
E^{j}_{i} = F_{i\alpha}F^{j\alpha} - \frac{1}{4}F_{\alpha\beta}F^{\alpha\beta}g^{j}_{i},
\end{equation} 
and
\begin{equation} 
\label{eq4}
\bar{p} = p - \xi v^{i}_{;i}.
\end{equation} 
Here $\rho$, $p$, $\bar{p}$ , $F^{j}_{i}$ and $\xi$ are the energy density, isotropic 
pressure, effective pressure, electromagnetic field tensor and bulk viscous coefficient 
respectively and $v^{i}$ is the flow vector satisfying the relation
\begin{equation} 
\label{eq5}
g_{ij} v^{i}v^{j} = - 1.
\end{equation} 
Here the semicolon represents a covariant differentiation. The coordinates are considered  
to be comoving so that $v^{1}$ = $0$ = $v^{2}$ = $v^{3}$ and $v^{4}$ = $\frac{1}{B}$. We 
consider the current to be flowing along the $z$-axis so that $F_{12}$ is the only 
non-vanishing component of $F_{ij}$. \\
The Einstein's field equations ( in gravitational units c = 1, G = 1 ) read as 
\begin{equation} 
\label{eq6} 
R^{j}_{i} - \frac{1}{2} R g^{j}_{i} + \Lambda g^{j}_{i} = - 8\pi T^{j}_{i},
\end{equation}  
for the line element (1) has been set up as
\begin{equation} 
\label{eq7}
8\pi B^{2}\left(\bar{p} + \frac{F^{2}_{12}}{2B^{4}}\right) = - \frac{B_{44}}{B} + 
\frac{B_{4}^{2}}{B^{2}} + \frac{B_{1}^{2}}{B^{2}} - \frac{C_{44}}{C} 
+ \frac{2 B_{1}C_{1}}{BC} - \Lambda B^{2},
\end{equation}  
\begin{equation} 
\label{eq8} 
8\pi B^{2}\left(\bar{p} + \frac{F^{2}_{12}}{2B^{4}}\right) =  
- \frac{B_{44}}{B} + \frac{B^{2}_{4}}{B^{2}} + \frac{B_{11}}{B}- \frac{B_{1}^{2}}{B^{2}} 
- \frac{C_{44}}{C} + \frac{C_{11}}{C} - \Lambda B^{2},
\end{equation}  
\begin{equation} 
\label{eq9} 
8\pi B^{2}\left(\bar{p} - \frac{F^{2}_{12}}{2B^{4}}\right) = 
- \frac{2B_{44}}{B} + \frac{B^{2}_{4}}{B^{2}} + \frac{2B_{11}}{B} - \frac{B^{2}_{1}}{B^{2}} 
- \Lambda B^{2},
\end{equation}  
\begin{equation} 
\label{eq10} 
8\pi B^{2}\left(\rho +  \frac{F^{2}_{12}}{2B^{4}}\right) = - \frac{B_{11}}{B} - \frac{C_{11}}{C} 
+ \frac{B^{2}_{1}}{B^{2}} + \frac{2B^{2}_{4}}{B^{2}} + \frac{2B_{4}C_{4}}{BC} + \Lambda B^{2},
\end{equation}  
\begin{equation} 
\label{eq11}
0 = \frac{B_{14}}{B} + \frac{C_{14}}{C} - \frac{2B_{1}B_{4}}{B^{2}} - \frac{B_{1}C_{4}}
{BC} - \frac{B_{4}C_{1}}{BC},
\end{equation}
where
\[
\bar{p} = p - \frac{\xi}{B}\left(\frac{2B_{4}}{B} + \frac{C_{4}}{C}\right).
\]
Here and in the following expressions the suffixes $1$ and $4$ at the symbols $B$, $C$, $f$
and $g$ denote differentiation with respect to $x$ and $t$ respectively.\\
%%%%%%%%%%%%%%%%%%%%%%%%%%%%%%%%%%%%%%%%%%%%%%%%%%%%%%%%%%%%%%%%%%%%%%%%%
%%%%%%%%%%%%%%%%%%%%%%%%%%  SECTION 3  %%%%%%%%%%%%%%%%%%%%%%%%%%%%%%%%%%
\section{Solution of the field equations}
From Equations (\ref{eq7}), (\ref{eq8}) and (\ref{eq9}), we have
\begin{equation} 
\label{eq12}
\frac{B_{11}}{B} + \frac{C_{11}}{C} - \frac{2B_{1}C_{1}}{BC} - \frac{2B^{2}_{1}}{B^{2}} = 0,
\end{equation} 
\begin{equation} 
\label{eq13}
\frac{8\pi F^{2}_{12}}{B^{2}} = \frac{B_{44}}{B} - \frac{B_{11}}{B} +  
\frac{C_{11}}{C} - \frac{C_{44}}{C}.
\end{equation}
Equations (\ref{eq7})-(\ref{eq11}) represent a system of five equations in seven 
unknowns $B$, $C$, $\rho$, $p$, $F_{12}$, $\Lambda$ and $\xi$. The research on exact
solutions is based on some physically reasonable restrictions used to simplify the 
Einstein equations. To get a determinate solution, we need two  extra conditions. \\
Let us consider that
\[
B = f(x) g(t),
\]
\begin{equation} 
\label{eq14}
C = h(x)k(t).
\end{equation}
Using Equation (\ref{eq14}) in Equation (\ref{eq11}), we obtain
\begin{equation} 
\label{eq15}
\frac{\frac{f_{1}}{f}}{\frac{h_{1}}{h}} = \frac{\frac{k_{4}}{k} - \frac{g_{4}}{g}}
{\frac{k_{4}}{k} + \frac{g_{4}}{g}} = K ~~(constant, say),
\end{equation}
which leads to
\begin{equation} 
\label{eq16}
\frac{f_{1}}{f} = K\frac{h_{1}}{h},
\end{equation}
and
\begin{equation} 
\label{eq17}
\frac{k_{4}}{k} - \frac{g_{4}}{g} = K\left(\frac{k_{4}}{k} + \frac{g_{4}}{g}\right).
\end{equation}
Using Equation (\ref{eq14}) in Equation (\ref{eq12}) leads to
\begin{equation} 
\label{eq18}
\frac{f_{11}}{f} - \frac{2f_{1}}{f}\left(\frac{h_{1}}{h} + \frac{f_{1}}{f}\right) + 
\frac{h_{11}}{h} = 0.
\end{equation}
Equations (\ref{eq16}) and (\ref{eq18}) give
\begin{equation} 
\label{eq19}
(K + 1)h h_{11} - (3K + K^{2})h^{2}_{1} = 0,
\end{equation}
which on integration leads to
\begin{equation} 
\label{eq20}
h = \left[\frac{K + 1}{(1 - 2K - K^{2})(\alpha x + \beta)}\right]^{\frac{(K + 1)}{(K^{2}
+ 2K -1)}},
\end{equation}
where $\alpha$ and $\beta$ are constants of integration. Integrating  Equations (\ref{eq16})
and (\ref{eq17}), we obtain 
\begin{equation} 
\label{eq21}
f = a h^{K},
\end{equation}
and
\begin{equation} 
\label{eq22}
k = \left(\frac{g}{b}\right)^{\frac{(1 + K)}{(1 - K)}}.
\end{equation}
respectively, where $a$ and $b$ are constants of integration. Hence from Equations (\ref{eq14}),
(\ref{eq20})-(\ref{eq22}), we obtain 
\begin{equation} 
\label{eq23}
B = ag \left[\frac{K + 1}{(1 - 2K - K^{2})(\alpha x + \beta)}\right]^{\frac{K(K + 1)}
{(K^{2} + 2K - 1)}},
\end{equation}
and
\begin{equation} 
\label{eq24}
 C = \left(\frac{g}{b}\right)^{\frac{(1 + K)}{(1 - K)}}\left[\frac{K + 1}{(1 - 2K - K^{2})
(\alpha x + \beta)}\right]^{\frac{(K + 1)}{(K^{2} + 2K -1)}}.
\end{equation}
After suitable transformation of coordinates and by taking $\alpha$ as unity without any
loss of generality the metric (\ref{eq1}) reduces to the form
\[
ds^{2} = M^{2} \left(\frac{1}{X^{2}}\right)^{\frac{K(K + 1)}{(K^{2} + 2K -1)}} g^{2}(T)
\left(dX^{2} - dT^{2} + dY^{2} \right) + 
\]
\begin{equation} 
\label{eq25}
\left(\frac{1}{X^{2}}\right)^{\frac{(K + 1)}
{(K^{2} + 2K -1)}} \left \{g^{2}(T)\right \}^{\frac{1 + K}{1 - K}}dZ^{2},
\end{equation}
where
\begin{equation} 
\label{eq26}
M = a\left[\frac{K + 1}{(1 - 2K - K^{2})}\right]^{\frac{K(K + 1)}{(K^{2}
+ 2K -1)}}.
\end{equation}
There is a lot of known solutions to the Einstein field equations but  Equation (\ref{eq25})
is indeed a new one. The effective pressure $\bar{p}$ and density $\rho$ for the model 
(\ref{eq25}) are given by
$$
8\pi \bar{p} = - \Lambda +  \frac{1}{g^{2} M^{2} X^{\frac{2K(K + 1)}{(1 - 2K - K^{2})}}}\times
$$
\begin{equation} 
\label{eq27} 
\left[\frac{2K^{2}(K + 1)(K + 2)}{(1 - 2K - K^{2})^{2}X^{2}} - \frac{(2 - K)}{(1 - K)}\frac{g_{44}}
{g} + \frac{(1 - 3K)}{(1 - K)^{2}}\frac{g^{2}_{4}}{g^{2}}\right],
\end{equation}
$$
8\pi \rho  = \Lambda + \frac{1}{g^{2}M^{2}X^{\frac{2K(K + 1)}{(1 - 2K - K^{2})}}}\times
$$
\begin{equation} 
\label{eq28}
\left[- \frac{2K(K + 1)(K + 2)}{(1 - 2K - K^{2})^{2}X^{2}} - \frac{1}{(1 - K)}\frac{g_{44}}{g} + 
\frac{(4 -5K -K^{2})}{(1- K)^{2}}\frac{g^{2}_{4}}{g^{2}}\right].
\end{equation}
For the specification of $\xi$, we assume that the fluid obeys an equation of state of the form
\begin{equation} 
\label{eq29}
p = \gamma \rho,
\end{equation}
where $\gamma(0 \leq \gamma \leq 1)$ is a constant.\\
Thus, given $\xi(t)$ we can solve for the cosmological parameters. In most of the 
investigations involving bulk viscosity is assumed to be a simple power function of 
the energy density\cite{ref83}$-$\cite{ref85}. 
\begin{equation}
\label{eq30}
\xi(t) = \xi_{0} \rho^{n},
\end{equation}
where $\xi_{0}$ and $n$ are constants. If $n = 1$, Equation (\ref{eq26}) may correspond
to a radiative fluid\cite{ref86}. However, more realistic models\cite{ref87} are 
based on $n$ lying in the regime $0 \leq n \leq \frac{1}{2}$. \\
On using (\ref{eq30}) in (\ref{eq27}), we obtain
$$
8\pi (p - \xi_{0}\rho^{n}\theta) = - \Lambda + \frac{1}{g^{2} M^{2} X^{\frac{2K(K + 1)}
{(1 - 2K - K^{2})}}}\times 
$$
\begin{equation} 
\label{eq31}
\left[\frac{2K^{2}(K + 1)(K + 2)}{(1 - 2K - K^{2})^{2}X^{2}} - \frac{(2 - K)}{(1 - K)}\frac{g_{44}}{g} 
+ \frac{(1 - 3K)}{(1- K)^{2}}\frac{g^{2}_{4}}{g^{2}}\right],
\end{equation}
where $\theta$ is the scalar of expansion calculated for the flow vector $v^{i}$ and is given by
\begin{equation} 
\label{eq32}
\theta = \frac{g_{4}}{g^{2} \sqrt{M} X^{\frac{K(K + 1)}{(1 - 2K - K^{2})}}}\left(\frac{K - 3}{K - 1}\right). 
\end{equation}
%%%%%%%%%%%%%%%%%%%%%%%%%%%%%%%%%%%%%%%%%%%%%%%%%%%%%%%%%%%%%%%%%%%%%%%%%%%%%%%%%
%%%%%%%%%%%%%%%%%%%  SUBSECTION 3.1  %%%%%%%%%%%%%%%%%%%%%%%%%%%%%%%%%%%%%%%%%%
\subsection {Model I: ~ ~ ~ $(\xi = \xi_{0})$}
When $n = 0$, Equation (\ref{eq30}) reduces to $\xi = \xi_{0}$. With the use of 
Equations (\ref{eq28}), (\ref{eq29}) and (\ref{eq32}), Equation (\ref{eq31}) reduces to
\[
8\pi (1 + \gamma) \rho = \frac{8\pi \xi_{0}g_{4}}{g^{2}\sqrt{M} X^{\frac{K(K + 1)}{(1 - 2K - K^{2})}}}
\left(\frac{K - 3}{K - 1}\right)  +
\]
\begin{equation} 
\label{eq33}
 \frac{1}{g^{2} M^{2} X^{\frac{2K(K + 1)}{(1 - 2K - K^{2})}}} 
\left[\frac{2K(K^{2} - 1)(K + 2)}{(1 - 2K - K^{2})^{2}X^{2}} - \frac{(3 - K)}{(1 - K)}\frac{g_{44}}{g} 
+ \frac{(5 - 8K - K^{2})}{(1- K)^{2}}\frac{g^{2}_{4}}{g^{2}}\right].
\end{equation} 
Eliminating $\rho(t)$ between (\ref{eq28}) and (\ref{eq33}), we get
$$
(1 + \gamma)\Lambda =  \frac{8\pi \xi_{0}g_{4}}{g^{2}\sqrt{M} X^{\frac{K(K + 1)}{(1 - 2K - K^{2})}}}
\left(\frac{K - 3}{K - 1}\right) + \frac{1}{g^{2} M^{2} X^{\frac{2K(K + 1)}{(1 - 2K - K^{2})}}}\times 
$$
\begin{equation} 
\label{eq34}
\left[\frac{2K(K + 1)(K + 2)(K + \gamma)}{(1 - 2K - K^{2})^{2}X^{2}}- \frac{(2 - K - \gamma)}
{(1 - K)}\frac{g_{44}}{g} + \frac{\left \{1 - 3K -(4 - 5K + K^{2})\gamma\right \}}{(1 - K^{2})}
\frac{g^{2}_{4}}{g^{2}}\right]. 
\end{equation}
%%%%%%%%%%%%%%%%%%%%%%%%%%%%%%%%%%%%%%%%%%%%%%%%%%%%%%%%%%%%%%%%%%%%%%%%%%%%%%%%%
%%%%%%%%%%%%%%%%%%%  SUBSECTION 3.2  %%%%%%%%%%%%%%%%%%%%%%%%%%%%%%%%%%%%%%%%%%
\subsection {Model II: ~ ~ ~ $(\xi = \xi_{0}\rho)$}
When $n = 1$, Equation (\ref{eq30}) reduces to $\xi = \xi_{0} \rho$. With the use of 
(\ref{eq28}), (\ref{eq29}) and (\ref{eq32}), Equation (\ref{eq31}) reduces to
\[
8\pi \left[1 + \gamma - \frac{\xi_{0}g_{4}}{g^{2}\sqrt{M} X^{\frac{K(K + 1)}{(1 - 2K - K^{2})}}}
\left(\frac{K - 3}{K - 1}\right)\right] \rho = 
\]
\begin{equation} 
\label{eq35}
\frac{1}{g^{2} M^{2} X^{\frac{2K(K + 1)}{(1 - 2K - K^{2})}}} 
\left[\frac{2K(K^{2} - 1)(K + 2)}{(1 - 2K - K^{2})^{2}X^{2}} - \frac{(3 - K)}{(1 - K)}\frac{g_{44}}{g} 
+ \frac{(5 - 8K - K^{2})}{(1- K)^{2}}\frac{g^{2}_{4}}{g^{2}}\right].
\end{equation}
Eliminating $\rho(t)$ between (\ref{eq28}) and (\ref{eq35}), we get
$$
\left[1 + \gamma - \frac{\xi_{0}g_{4}}{g^{2}\sqrt{M} X^{\frac{K(K + 1)}
{(1 - 2K - K^{2})}}}\left(\frac{K - 3}{K - 1}\right)\right] \Lambda =\frac{1}{g^{2} M^{2} X^{\frac{2K(K + 1)}
{(1 - 2K - K^{2})}}}\times 
$$
$$
\left[\frac{2K(K + 1)(K + 2) (K + \gamma)}{(1 - 2K - K^{2})^{2}X^{2}} - \frac{1 - \gamma(2 - K)}
{(1 - K)}\frac{g_{44}}{g} + \frac{(4 - 5 K - K^{2}) - (1 - 3K)\gamma}{(1 - K)^{2}}
\frac{g^{2}_{4}}{g^{2}} \right]
$$
$$
+ \frac{\xi_{0}g_{4}}{g^{4} M^{\frac{3}{2}} X^{\frac{3K(K + 1)}{(1 - 2K - K^{2})}}}
\left(\frac{K - 3}{K - 1}\right) \times
$$
\begin{equation} 
\label{eq36}
\left[- \frac{2K(K + 1)(K + 2)}{(1 - 2K - K^{2})^{2}X^{2}} - \frac{(2 - K)}{(1 - K)}\frac{g_{44}}{g} 
+ \frac{(1 - 3K)}{(1- K)^{2}}\frac{g^{2}_{4}}{g^{2}}\right].
\end{equation}
   In spite of homogeneity at large scale our universe is inhomogeneous at small
scales, so physical quantities having position dependent are more natural in
our observable universe if we do not go to super high scale. This result shows 
this kind of physical importance. In recent times the $\Lambda$-term has interested
theoreticians and observers for various reasons. The nontrivial role of the vacuum in 
the early universe generate a $\Lambda$-term that leads to inflationary phase. Observationally
this term provides an additional parameter to accommodate conflicting data on the values of the 
Hubble constant, the deceleration parameter, the density parameter and the age of the universe
(for example, see the references \cite{ref89},\cite{ref90}). Assuming that $\Lambda$ owes its
origin to vacuum interactions, as suggested in particular by Sakharov \cite{ref91} it follows 
that it would in general be a function of space and time coordinates, rather than a strict
constant. In a homogeneous universe $\Lambda$ will be at most time dependent
\cite{ref92}. In our case this approach can generate $\Lambda$ that varies both with space 
and time. In considering the nature of local massive objects, however, the space dependence
of $\Lambda$ cannot be ignored. For details discussion, the readers are advised to see
the references (Narlikar, Pecker and Vigier \cite{ref93}, Ray {\it et al.} \cite{ref94}).  

   The effect of bulk viscosity is to introduce a change in the perfect fluid model. We also 
observe here that the condition of Murphy \cite{ref88} about the absence of a big bang type 
of singularity in the finite past in models with bulk viscous fluid is, in general, not true. 
We have freedom of choosing the function $g(T)$ so that to give a physical behaviour 
of above parameters. As a matter of fact, there are multiple choices, for example,
$g(T) = c^{2} + d^{2} t^{2}, c^{2} + e^{- d ~ ~ t^{2}}, c^{2} + d^{2} \cos \omega t, 
c^{2} > d^{2}$, where $c$ and $d$ are some real constants. From Equations (\ref{eq34}) 
and (\ref{eq36}), we observe that the cosmological constant is a decreasing function 
of time and it approaches a small positive value at late times under some suitable 
conditions which explains the small value of $\Lambda$ at present. 
%%%%%%%%%%%%%%%%%%%%%%%%%%%%%%%%%%%%%%%%%%%%%%%%%%%%%%%%%%%%%%%%%%%%%%%%%%%%%%%%%
%%%%%%%%%%%%%%%%%%%  SUBSECTION 3.3  %%%%%%%%%%%%%%%%%%%%%%%%%%%%%%%%%%%%%%%%%%
\subsection {Some physical aspects of the models}
We shall now give the expressions for kinematical quantities and the components of conformal 
curvature tensor. With regard to the kinematical properties of the velocity vector $v^{i}$ in
the metric (\ref{eq25}), a straightforward calculation leads to the following expression for 
the shear of the fluid: 
\begin{equation} 
\label{eq37}
\sigma^{2} = \frac{4K^{2}{g_{4}}^{2}}{3\sqrt{M}g^{4}(1 - K)^{2}X^{\frac{2K(K + 1)}
{(1 - 2K - K^{2})}}}.
\end{equation}
The rotation $\omega$ and acceleration are identically zero. The expansion scalar $\theta$
has already been given by (\ref{eq32}). The non-vanishing physical components of the conformal
curvature tensor are given by
$$
C_{(1212)} = - C_{(3434)} = \frac{1}{6Mg^{2}X^{\frac{2K(K + 1)}{(1 - 2K - K^{2})}}}\times
$$
\begin{equation} 
\label{eq38}
\left[\frac{2K(K + 1)(K - 1)}
{(1 - 2K - K^{2})^{2}X^{2}} - \frac{4K}{(1 - K)}\frac{g_{44}}{g} + \frac{4K(1 - 3K)}{(1 - K)^{2}}
\frac{g_{4}^{2}}{g^{2}}\right],
\end{equation}
$$
C_{(1313)} = - C_{(2424)} = \frac{1}{6Mg^{2}X^{\frac{2K(K + 1)}{(1 - 2K - K^{2})}}}\times
$$
\begin{equation} 
\label{eq39}
\left[\frac{4K(K + 1)(1 - K)}
{(1 - 2K - K^{2})^{2}X^{2}} + \frac{2K}{(1 - K)}\frac{g_{44}}{g} + \frac{2K(3K - 1)}{(1 - K)^{2}}
\frac{g_{4}^{2}}{g^{2}}\right],
\end{equation}
\begin{equation} 
\label{eq40}
C_{(1224)} = - C_{(1334)} = \frac{K(K + 1)}{M(1 - 2K - K^{2})X^{\frac{2K(K + 1)}{(1 - 2K - K^{2})}}}
\frac{g_{4}}{g}, 
\end{equation}
\begin{equation} 
\label{eq41}
C_{(2323)} = - C_{(1414)} = \frac{1}{3Mg^{2}X^{\frac{2K(K + 1)}{(1 - 2K - K^{2})}}}\left[\frac{K(K + 1)^{2}
(1 - K)}{(1 - 2K - K^{2})^{2}X^{2}} -  \frac{2K}{(1 - K)}\frac{g_{4}^{2}}{g^{2}}\right].
\end{equation}
The non-vanishing component $F_{12}$ of the electromagnetic field tensor and $J^{2}$, the component
of charge current density, are given by
\begin{equation} 
\label{eq42}
F_{12} = \frac{1}{2\sqrt{\pi}} \left[-\frac{K}{(1 - K)}\frac{g_{44}}{g} -
\frac{K(1 + K)}{(1 - K)^{2}}\frac{g^{2}_{4}}{g^{2}} - \frac{K(K +1)(K - 1)(K + 2)}
{(1 - 2K - K^{2})X^{2}}\right]^{\frac{1}{2}}B,
\end{equation}
$$
J^{2} = \frac{1}{M^{2}g^{4}X^{\frac{(1 + K)(1 + 3K)}{(1 - 2K - K^{2})}}}\times  
$$
\begin{equation} 
\label{eq43}
\left[\frac{\sqrt{\zeta}(1 - K^{2})}{(1 - 2K - K^{2})}X^{\frac{2K}{(1 - 2K - K^{2})}} + 
\frac{2K(K + 1)(K - 1)(K + 2)\zeta^{-\frac{1}{2}}}{(1 - 2K - K^{2})^{2}}X^{\frac{2(K^{2} 
+ 3K -1)}{(1 - 2K - K^{2})}}\right],
\end{equation}
where
\begin{equation} 
\label{eq44}
\zeta = \left[-\frac{2K}{(1 - K)}\frac{g_{44}}{g} - \frac{2K(1 + K)}{(1 - K)^{2}}
\frac{g^{2}_{4}}{g^{2}} - \frac{2K(K + 1)(K - 1)(K + 2)}{(1 - 2K - K^{2})^{2}X^{2}}\right].
\end{equation}
The models represent shearing, non-rotating and Petrov type I non-degenerate in general, 
in which the flow is geodetic. The expansion in the model stops when $K = 3$ but it will 
continue indefinitely when $K > 3$. Since $\lim_{T \rightarrow 0} \frac{\sigma}{\theta} 
\ne 0$, hence the models do not approach isotropy for large values of $T$.  
%%%%%%%%%%%%%%%%%%%%%%%%%%%%%%%%%%%%%%%%%%%%%%%%%%%%%%%%%%%%%%%%%%%%%%%%%
%%%%%%%%%%%%%%%%%%%%%%%%%%  SECTION 4  %%%%%%%%%%%%%%%%%%%%%%%%%%%%%%%%%%
\section{Other generated model}
In the metric (\ref{eq25}), the function $g(T)$ is indeterminate. To get the determinate 
value of $g$, we assume in Equation (\ref{eq15}) as
\begin{equation} 
\label{eq45}
\frac{k_{4}}{k} - \frac{g_{4}}{g} = r,
\end{equation}
\begin{equation} 
\label{eq46}
\frac{k_{4}}{k} + \frac{g_{4}}{g} = s,
\end{equation}
where $r$ and $s$ are constants. From Equations (\ref{eq45}) and (\ref{eq46}), 
we have derived 
\begin{equation} 
\label{eq47}
k = l e^{\frac{(r + s)}{2}T},
\end{equation}
\begin{equation} 
\label{eq48}
g = m e^{\frac{(r - s)}{2}T},
\end{equation}
where $l$ and $m$ are integrating constants. In this case the geometry of the universe 
(\ref{eq25}) reduces to the form
$$
ds^{2} = N^{2} \left(\frac{1}{X^{2}}\right)^{\frac{\kappa(\kappa + 1)}{(\kappa^{2} + 
2\kappa  - 1)}}e^{\frac{(s - r)}{2}T} (dX^{2} - dT^{2} + dY^{2}) 
$$
\begin{equation} 
\label{eq49}
+ \left(\frac{1}{X^{2}}\right)^{\frac{(\kappa + 1)}{(\kappa^{2} + 2\kappa  - 1)}}
e^{\frac{(r + s)}{2}T} dZ^{2},
\end{equation}
where
\begin{equation} 
\label{eq50}
N = a m \left[\frac{\kappa + 1}{(1 - 2\kappa - \kappa^{2})}\right]^{\frac{\kappa 
(\kappa + 1)}{(\kappa^{2} + 2 \kappa - 1)}} ~ ~ and ~~ \kappa = \frac{r}{s}. 
\end{equation}
The effective pressure $\bar{p}$ and density $\rho$ for the model (\ref{eq49}) 
are given by
\begin{equation} 
\label{eq51}
8 \pi \bar{p} = \frac{e^{\frac{(r - s)}{2}T}}{N^{2} X^{\frac{2 \kappa(\kappa + 1)}
{(1 - 2\kappa - \kappa^{2})}}}\left[\frac{2\kappa^{2}(\kappa + 1)(\kappa + 2)}{(1 - 2\kappa 
- \kappa^{2}) X^{2}} - \frac{1}{16}(r^{2} + s^{2})\right] - \Lambda,
\end{equation}
\begin{equation} 
\label{eq52}
8 \pi \rho = \frac{e^{\frac{(r - s)}{2}T}}{N^{2} X^{\frac{2 \kappa(\kappa + 1)}{(1 - 2\kappa
- \kappa^{2})}}}\left[\frac{- 2\kappa(\kappa + 1)(\kappa + 2)}{(1 - 2\kappa - \kappa^{2}) X^{2}}
+ \frac{1}{8}\{s(2s - r)\}\right] + \Lambda.
\end{equation}
On using Equations (\ref{eq4}) and (\ref{eq30}) in (\ref{eq51}), we obtain
\begin{equation} 
\label{eq53}
8 \pi (p - \xi_{0}\rho^{n} \theta)  = \frac{e^{\frac{(r - s)}{2}T}}{N^{2} X^{\frac{2 \kappa(\kappa
+ 1)}{(1 - 2\kappa - \kappa^{2})}}}\left[\frac{2\kappa^{2}(\kappa + 1)(\kappa + 2)}{(1 - 2\kappa 
- \kappa^{2}) X^{2}} - \frac{1}{16}(r^{2} + s^{2})\right] - \Lambda,
\end{equation}
where $\theta$ is the scalar expansion calculated for the flow vector $v^{i}$ and is given by
\begin{equation} 
\label{eq54}
\theta = \frac{(r + 3s) e^{\frac{(r - s)}{4}T}}{4\sqrt{N}X^{\frac{\kappa(\kappa + 1)}
{(1 - 2\kappa - \kappa^{2})}}}.
\end{equation}
%%%%%%%%%%%%%%%%%%%%%%%%%%%%%%%%%%%%%%%%%%%%%%%%%%%%%%%%%%%%%%%%%%%%%%%%%%%%%%%%%
%%%%%%%%%%%%%%%%%%%  SUBSECTION 4.1  %%%%%%%%%%%%%%%%%%%%%%%%%%%%%%%%%%%%%%%%%%
\subsection {Model I: Solution for $(\xi = \xi_{0})$}
In this case, using  Eqs. (\ref{eq52}), (\ref{eq29}) and (\ref{eq54}) in Eq. (\ref{eq53}),
we obtain
$$
8 \pi (1 + \gamma)\rho = \frac{8 \pi \xi_{0} (r + 3s)e^{\frac{(r - s)}{4} T}}
{4 \sqrt{N} X^{\frac{\kappa(\kappa + 1)}{(1 - 2\kappa - \kappa^{2})}}} +
$$
\begin{equation} 
\label{eq55}
\frac{e^{\frac{(r - s)}{2}T}}{N^{2} X^{\frac{2 \kappa(\kappa + 1)}{(1 - 2\kappa
- \kappa^{2})}}}\left[\frac{2\kappa (\kappa + 1)(\kappa - 1)(\kappa + 2)}{(1 - 2\kappa -
\kappa^{2}) X^{2}} - \frac{1}{16} \left \{r^{2} + 2rs - 3s^{2}\right \} \right].
\end{equation}
Eliminating $\rho(t)$ between (\ref{eq52}) and (\ref{eq55}), we get
$$
(1 + \gamma)\Lambda  = \frac{8 \pi \xi_{0} (r + 3s)e^{\frac{(r - s)}{4} T}}
{4 \sqrt{N} X^{\frac{\kappa(\kappa + 1)}{(1 - 2\kappa - \kappa^{2})}}} +
$$
\begin{equation} 
\label{eq56}
\frac{e^{\frac{(r - s)}{2}T}}{N^{2} X^{\frac{2 \kappa(\kappa + 1)}{(1 - 2\kappa
- \kappa^{2})}}}\left[\frac{2\kappa (\kappa + 1)(\kappa + 2)(\kappa + \gamma)}
{(1 - 2\kappa - \kappa^{2}) X^{2}} - \frac{1}{16} \left \{r^{2} + 2rs (2 + \gamma) 
- s^{2}(7 + 4\gamma)\right \} \right].
\end{equation}
%%%%%%%%%%%%%%%%%%%%%%%%%%%%%%%%%%%%%%%%%%%%%%%%%%%%%%%%%%%%%%%%%%%%%%%%%%%%%%%%%
%%%%%%%%%%%%%%%%%%%  SUBSECTION 4.2  %%%%%%%%%%%%%%%%%%%%%%%%%%%%%%%%%%%%%%%%%%
\subsection {Model II: Solution for $(\xi = \xi_{0}\rho)$}
In this case with the use of Eqs. (\ref{eq52}), (\ref{eq29}) and (\ref{eq54}) in Eq. 
(\ref{eq53}) reduces to
$$
8 \pi \left[1 + \gamma  - \frac{\xi_{0} (r + 3s)e^{\frac{(r - s)}{4} T}}
{4 \sqrt{N} X^{\frac{\kappa(\kappa + 1)}{(1 - 2\kappa - \kappa^{2})}}}\right]\rho = 
$$
\begin{equation} 
\label{eq57}
\frac{e^{\frac{(r - s)}{2}T}}{N^{2} X^{\frac{2 \kappa(\kappa + 1)}{(1 - 2\kappa
- \kappa^{2})}}}\left[\frac{2\kappa (\kappa + 1)(\kappa - 1)(\kappa + 2)}{(1 - 2\kappa - \kappa^{2})
X^{2}} - \frac{1}{16} \left \{r^{2} + 2rs - 3s^{2}\right \} \right].
\end{equation}
Eliminating $\rho(t)$ between (\ref{eq52}) and (\ref{eq57}), we get
$$
\left[1 + \gamma  - \frac{\xi_{0} (r + 3s)e^{\frac{(r - s)}{4} T}}
{4 \sqrt{N} X^{\frac{\kappa(\kappa + 1)}{(1 - 2\kappa - \kappa^{2})}}}\right]\Lambda = 
\frac{e^{\frac{(r - s)}{2}T}}{N^{2} X^{\frac{2 \kappa(\kappa + 1)}{(1 - 2\kappa
- \kappa^{2})}}}\Biggl[\frac{2\kappa^{2} (\kappa + 1)(\kappa + 2)}{(1 - 2\kappa - \kappa^{2})
X^{2}} - 
$$
\begin{equation} 
\label{eq58}
\frac{1}{16}(r^{2} + s^{2}) + \left \{\frac{2\kappa(\kappa +1)(\kappa + 2)}
{(1 - 2\kappa - \kappa^{2})X^{2}} + \frac{1}{8}(r - 2s)s \right \}\times \left \{\gamma - 
\frac{(r + 3s)\xi_{0}e^{\frac{(r - s)}{4}T}}
{4\sqrt{N} X^{\frac{\kappa(\kappa + 1)}{(1 - 2\kappa - \kappa^{2})}}}\right \}\Biggr].
\end{equation}
From Equations (\ref{eq56}) and (\ref{eq58}), we observe that cosmological constant $\Lambda$ 
may be positive or negative under specific conditions. A negative cosmological constant adds 
to the attractive gravity of matter, therefore, universe with a negative cosmological constant 
are invariably doomed to recollapse. A positive cosmological constant resists the attractive 
gravity of matter due to its negative pressure. For most universe, the positive cosmological 
constant eventually dominates over the attraction of matter and drives the universe to expands 
exponentially. 
%%%%%%%%%%%%%%%%%%%%%%%%%%%%%%%%%%%%%%%%%%%%%%%%%%%%%%%%%%%%%%%%%%%%%%%%%%%%%%%%%
%%%%%%%%%%%%%%%%%%%  SUBSECTION 4.3  %%%%%%%%%%%%%%%%%%%%%%%%%%%%%%%%%%%%%%%%%%
\subsection {Some physical aspects of the models}
The coefficient of shear $\sigma$, non-vanishing physical components of conformal curvature 
tensor $C_{(ijkl)}$, non-vanishing component of electromagnetic field tensor $F_{ij}$ and component of 
charge density $J^{2}$ for the model (\ref{eq49}) are given as:
\begin{equation} 
\label{eq59}
\sigma^{2} = \frac{r^{2} e^{\frac{(r - s)}{2}T}}{8N X^{\frac{2\kappa(\kappa + 1)}
{(1 - 2\kappa - \kappa^{2})}}},
\end{equation}
\begin{equation} 
\label{eq60}
C_{(1212)} = - C_{(3434)} =  \frac{ e^{\frac{(r - s)}{2}T}}{6N X^{\frac{2\kappa(\kappa + 1)}
{(1 - 2\kappa - \kappa^{2})}}}\left[\frac{2\kappa(\kappa + 1)(\kappa - 1)}{(1 - 2\kappa - 
\kappa^{2})^{2}X^{2}} - \frac{r^{2}}{2}\right],
\end{equation}
\begin{equation} 
\label{eq61}
C_{(1313)} = - C_{(2424)} =  \frac{ e^{\frac{(r - s)}{2}T}}{6N X^{\frac{2\kappa(\kappa + 1)}
{(1 - 2\kappa - \kappa^{2})}}}\left[- \frac{4\kappa(\kappa + 1)(\kappa - 1)}{(1 - 2\kappa - 
\kappa^{2})^{2}X^{2}} + \frac{r^{2}}{4}\right],
\end{equation}
\begin{equation} 
\label{eq62}
C_{(1224)} = - C_{(1334)} =  \frac{r  e^{\frac{(r - s)}{2}T}}{4N X^{\frac{2\kappa(\kappa
+ 1)}{(1 - 2\kappa - \kappa^{2})}}},
\end{equation}
\begin{equation} 
\label{eq63}
C_{(2323)} = - C_{(1414)} =  \frac{ e^{\frac{(r - s)}{2}T}}{3N X^{\frac{2\kappa(\kappa + 1)}
{(1 - 2\kappa - \kappa^{2})}}}\left[- \frac{\kappa(\kappa + 1)^{2}(\kappa - 1)}{(1 - 2\kappa - 
\kappa^{2})^{2}X^{2}} + \frac{1}{8}\left \{r(r - s)\right \}\right],
\end{equation}
\begin{equation} 
\label{eq64}
F_{12} = \frac{1}{\sqrt{8\pi}}\left[\frac{2\kappa(\kappa + 1)(1 - \kappa)(\kappa + 2)}{(1 - 2\kappa
- \kappa^{2}) X^{2}} - \frac{rs}{4}\right]^{\frac{1}{2}} B,
\end{equation}
$$
J^{2} = - \frac{e^{\frac{(r - s)}{2}T}}{N^{2} X^{\frac{(\kappa + 1)(3\kappa + 1)}
{(1 - 2\kappa - \kappa^{2})}}}\times
$$
\begin{equation} 
\label{eq65}
\left[\frac{(1 - \kappa^{2})}{(1 - 2\kappa - \kappa^{2})}\psi^{\frac{1}{2}}
X^{\frac{2\kappa}{(1 - 2\kappa - \kappa^{2})}} +
 \frac{2\kappa(\kappa + 1)(\kappa - 1)(\kappa + 2)}
{(1 - 2\kappa - \kappa^{2})^{2}} \psi^{-\frac{1}{2}} X^{\frac{2(\kappa^{2} + 3\kappa -1)}{(1 - 2\kappa
- \kappa^{2})}}\right],
\end{equation}
where
\begin{equation} 
\label{eq66}
\psi = \left[\frac{2\kappa(\kappa + 1)(1 - \kappa)(\kappa + 2)}{(1 - 2\kappa
- \kappa^{2}) X^{2}} - \frac{rs}{4}\right].
\end{equation}
The rotation $\omega$ is identically zero. The models start expanding at $T = 0$ and continue
 till $T = \infty$ for $N > 0$. The expansion stops when $T = -\infty$ or $r = - 3s$. The models 
represent shearing, non-rotating and Petrov type I non-degenerate in general.
%%%%%%%%%%%%%%%%%%%%%  SECTION 5  %%%%%%%%%%%%%%%%%%%%%%%%%%%%%%%%%%%%%%%%%%
\section {Conclusions}
We have obtained a new class of plane-symmetric inhomogeneous cosmological models of electromagnetic
bulk viscous fluid as the source of matter. Generally the models represent expanding, shearing,
non-rotating and Petrov type I non-degenerate universe in which the flow vector is geodetic. In all 
these models, we observe that they do not approach isotropy for large values of time.\\

          The cosmological constants in all models given in Sections $3$ are decreasing 
functions of time and they all approach a small value at late times.
The values of cosmological ``constant'' for these models are found to be small and positive which 
are supported by the results from recent supernova Ia observations recently obtained by 
the High-Z Supernova Team and Supernova Cosmological Project ( Garnavich {\it et al.}\cite{ref30};
Perlmutter {\it et al.}\cite{ref27}; Riess {\it et al.}\cite{ref28}; Schmidt {\it et al.}
\cite{ref33}). 
\section*{Acknowledgements}
One of the authors (A. Pradhan) would like to thank the Inter-University
Centre for Astronomy and Astrophysics, Pune, India for providing facility under Associateship 
Programme where part of this work was carried out. Authors thank the anonymous referee for many
helpful comments which helped in improving the presentation of this paper. Authors also thank S. K. 
Srivastava and Saibal Ray for useful discussions.
%\newline
%\newline

\end{document}